\documentstyle[pra,multicol,aps]{revtex}
\begin{document}

\setlength{\textwidth}{150mm}
\setlength{\textheight}{240mm}
\setlength{\parskip}{2mm}

\input{epsf.tex}
\epsfverbosetrue

\title{Induced Coherence and Stable Soliton Spiraling}

\author{Alexander V.~Buryak$^{1,2}$, Yuri S. Kivshar$^{2}$,
Ming-feng Shih$^{3,4}$, and Mordechai Segev$^{3,5}$}

\address{$^1$School of Mathematics and Statistics,
Australian Defence Force Academy, Canberra ACT 2600, Australia\\ $^2$ 
Optical Sciences Centre, Australian National
University,  Canberra ACT 0200, Australia\\ 
$^3$ Department of Electrical Engineering, Princeton
University, Princeton, NJ 08544 \\
$^4$ Department of Physics, National Taiwan University, Taipei,
Taiwan, Republic of China\\
$^5$ Physics Department, Technion-Israel Institute of
Technology, Haifa 32000, Israel}

\maketitle
\begin{abstract}
We develop a theory of soliton spiraling in a bulk nonlinear
medium and reveal a new physical mechanism:  periodic power
exchange via induced coherence, which can lead to
stable spiraling  and the formation of dynamical two-soliton states. Our 
theory not only explains earlier observations, but
provides a number of predictions which are also verified experimentally. 
Finally, we show theoretically and experimentally that soliton spiraling 
can be controled by the degree of mutual initial coherence.
\end{abstract}


\vspace*{-1.0 cm}
\begin{multicols}{2}
\narrowtext

Self-guided optical beams (or {\em spatial solitons}) have attracted
substantial research interest in the last three decades
\cite{barth}. Although interactions between two-dimensional (2D) solitons 
in Kerr and non-Kerr media have been studied extensively, 
only the recent discoveries of stable three-dimensional (3D) solitons in 
different nonlinear bulk media \cite{soliton_3D} initiated
an experimental study of fully 3D interactions of solitary waves.  
Recently, experiments demonstrating non-planar interaction and spiraling 
of spatial solitons in a photorefractive medium have been
reported \cite{Ming}.  However, in spite of earlier predictions of
 non-planar soliton interactions \cite{snyder}, the experimental results 
\cite{Ming} have not been explained
theoretically so far. Also, it has been shown \cite{Vika}  that
{\em coherently interacting} solitons do not allow {\em any stable
spiraling}, in sharp contrast with the experimental observations 
\cite{Ming}. The fundamental question remains: {\em Is soliton spiraling 
possible at all as a stable dynamical regime of soliton interaction?}

In this Letter we develop, for the first time to our knowledge,
a theory of soliton spiraling in a photorefractive-type nonlinear
bulk medium.  We derive an analytical model describing stable
soliton spiraling and predict a number of new effects in
soliton interactions, such as {\em an induced coherence} and
{\em control over 3D interactions}, which we verify here
experimentally, using experimental setup similar to that reported earlier 
\cite{Ming}. Importantly, our analytical model and numerical simulations 
show that interacting-spiraling solitons conserve angular momentum. We 
believe that this result
is a core foundation for future research on 3D soliton control, resembling 
the conservation of linear momentum in the interaction of more 
conventional (1+1)-dimensional solitons \cite{Kruskal}.

We consider incoherent beam interaction in an isotropic saturable 
nonlinear medium described by two coupled normalized nonlinear
Schr\"odinger (NLS) equations
\begin{equation}
\label{norm_eqs}
\begin{array}{l} {\displaystyle i \frac{\partial u}{\partial z}
+ \nabla_\perp^2 u - \frac{u}{1+|u|^2 + |w|^2} =0,} \\*[9pt]
{\displaystyle i \frac{\partial w}{\partial z} + \nabla_\perp^2
w - \frac{w}{1+|u|^2 + |w|^2} = 0,}
\end{array}
\end{equation}
where $u$ and $w$ are the beam envelopes, $z$ is the propagation distance; 
$\nabla_\perp^2 \equiv \partial^2/\partial x^2 + \partial^2/\partial y^2$  
accounts for the diffraction in the transverse ($x, y)$ plane. This 
system, in the 2D case (i.e., for $\nabla_\perp^2 \equiv 
\partial^2/\partial x^2$), gives rise to
incoherently-coupled soliton pairs \cite{Christol} and to
incoherent collisions \cite{Shihl} which have both been demonstrated with 
photorefractive screening solitons \cite{one_dim}.

We look for solitary waves of Eqs. (\ref{norm_eqs})  in the form $u = U(r) 
\exp{(i \beta_{u} z)}$, $w = W(r) \exp{(i \beta_{w}
z)}$, where the envelopes $U$ and $W$ satisfy the equations
\begin{equation}
\label{stat_eqs}
\begin{array}{l}
{\displaystyle \frac{d^2 U}{d r^2} + \frac{1}{r}\frac{d U}{d r}
-\beta_u U - \frac{U}{1+U^2 + W^2} = 0,} \\*[9pt]
{\displaystyle \frac{d^2 W}{d r^2} + \frac{1}{r}\frac{d W}{d r}
-\beta_w W - \frac{W}{1+U^2 + W^2} = 0.}
\end{array}
\end{equation}
Here $r \equiv \sqrt{x^2+y^2}$ is the radial coordinate, and 
$\beta_u$ and $\beta_w$ are nonlinearity-induced shifts of the propagation 
constants. System (\ref{stat_eqs}) has two families of soliton solutions: 
$U = G_u(\beta_u, r)$, $W = 0$ and $U = 0$, $W = G_w(\beta_w, r)$, which 
can be found numerically by solving the equation $G^{\prime 
\prime}_{\alpha} + G^{\prime}_{\alpha}/r - {\beta}_{\alpha} {G}_{\alpha} - 
{G}_{\alpha}/(1+G^{2}_{\alpha}) = 0$, where $\alpha = \{u,w\}$. These 
solutions can be characterised by the soliton powers  $P(\beta_\alpha) 
\equiv 2 \pi \int_{0}^{\infty} G_\alpha^2(\beta_\alpha, r) \; r dr$.

In addition to the one-component solitons, at $\beta_u = \beta_v \equiv 
\beta$ there exists a family of two-component ({\em vector}) solitons 
defined as: $U = G(\beta, r) \cos{\theta}, \; W = G(\beta, r) 
\sin{\theta}$, where the variable $\theta$ characterises a power 
distribution between the components.  Moving solitons of Eqs.  
(\ref{norm_eqs}) can be obtained by a well-known gauge transformation.

To  study the soliton collisions analytically, we employ a Lagrangian
formalism \cite{Anderson,Malomed}. Equations (\ref{norm_eqs})
can be obtained from the Lagrangian density:
${\cal{L}} = (i/2) (u^* u_z - u u_z^*) - (|u_x|^2 + |u_y|^2) + (i/2)(w^* 
w_z - w w_z^*) - (|w_x|^2 + |w_y|^2) - \ln (1 + |u|^2 + |w|^2)$.
Now we consider the interaction between two spatial solitons $(u_1,w_1)$ 
and $(u_2,w_2)$,  taking $u = u_1 + u_2$, $w = w_1 + w_2$ and introducing 
the following free parameters $(j =1,2)$: the positions of soliton centers 
($x_j$,$y_j$), and the common and relative phases of the soliton 
components $u_j$ and $v_j$, which we denote by $\phi_{j}$ and $\psi_{j}$,  
respectively. Following \cite{Malomed}, we assume that the soliton 
parameters vary slowly in $z$, and integrate the Lagrangian density over 
$x$ and $y$. After this averaging  procedure, we reduce the number of 
equations by using the conservation of the angular  momentum,  ${\cal M} = 
s V_{0} P/4$, where $s$ is the impact parameter (defined as the minimum 
distance between the trajectories of non-interacting solitons), and 
$V_0 \equiv d{R}_0/dz$ is the initial value of the soliton relative 
velocity (see Ref. \cite{Vika}).

The  averaged Lagrangian can be presented as
$L = L_1+ L_2 - U_{\rm int}$, where the first two terms are
the individual contributions of the vector solitons,  and 
the third term corresponds to an effective interaction potential 
given by
\begin{eqnarray}
\label{Inter}
{U_{\rm int} = M_{R} s^2 V_0^2/(2 R^2) -  U_{\rm incoh}(R) - U_{\rm 
coh}(R) \times} \nonumber \\ {\left[ \cos{\theta_{-}}  \cos{\phi_{-}} 
\cos{\left(\frac{\psi_{-}}{2}\right)} + \cos{\theta_{+}} \sin{\phi_{-}} 
\sin{\left(\frac{\psi_{-}}{2}\right)}\right]},
\end{eqnarray}
where $M_{R} \equiv P/2$, $\theta_{\pm} \equiv \theta_{2} \pm \theta_{1}$, 
$\phi_{\pm} \equiv \phi_2 \pm \phi_1$, $\psi_{\pm} \equiv \psi_2 \pm 
\psi_1$, and $R \equiv \sqrt{(x_2-x_1)^{2}+(y_2-y_1)^{2}}$  is the 
relative distance between the interacting solitons. The functions $U_{\rm 
incoh}$ and $U_{\rm coh}$ are expressed in terms of the soliton overlap 
integrals, $U_{\rm coh} = 2 \int \int_{-\infty}^{\infty}
[G^{3}_{1}G_{2}/(1+G^{2}_{1}) + G^{3}_{2}G_{1}/(1+G^{2}_{2})] dx \; dy$, 
$U_{\rm incoh} = \int \int_{-\infty}^{\infty} [G^{2}_{1} 
G^{2}_{2}/(1+G^{2}_{1})  + G^{2}_{2} G^{2}_{1} / (1+G^{2}_{2})] dx \; 
dy$.  The first term in Eq. (\ref{Inter}) describes a {\em centrifugal 
force} (which is always repulsive), the second -- {\em incoherent 
attraction}, and the third -- {\em coherent interaction}.  When $R$ is 
large enough,  the soliton interaction is determined by the tail
asymptotics  $G(r) \sim \exp{[-\sqrt{(1+\beta)} r]}/\sqrt{r}$,
that yields, $U_{\rm coh}(R) \sim \exp{[-\sqrt{(1+\beta)} R]}/\sqrt{R}$, 
$U_{\rm incoh}(R) \sim \exp{[-2\sqrt{(1+\beta)} R]}/R$ and $U_{\rm coh} 
\gg U_{\rm incoh}$.  For smaller $R$, although $U_{\rm coh} > U_{\rm 
incoh}$,
$U_{\rm incoh}$ is also important.

The average Lagrangian generates the following equations,
\begin{eqnarray}
\label{full_system}
{ \hspace*{-10mm} M_{R} \ddot{R} + \frac{\partial U_{\rm int}}{\partial
R} = 0, \;\;\;\;  M_{\phi}
 \ddot{\phi}_{-} +   \frac{\partial
U_{\rm int}}{\partial \phi_{-}} = 0,} \\
{M_{\psi} \dot{\psi}_{-} - \cos{\theta_{+}}
\sin{\theta_{-}} \frac{\partial U_{\rm int}}{\partial \theta_{+}}+
\sin{\theta_{+}} \cos{\theta_{-}}
\frac{\partial U_{\rm int}}{\partial \theta_{-}} = 0, }\nonumber \\ 
{M_{\theta} \dot{\theta}_{\mp} \mp \frac{1}{2} \sin{\theta_{\pm}}
\cos{\theta_{\pm}} \frac{\partial U_{\rm int}}{\partial \phi_{-}} \mp  
\sin{\theta_{\pm}} \cos{\theta_{\mp}}
\frac{\partial U_{\rm int}}{\partial \psi_{-}} = 0,}  \nonumber
\end{eqnarray}
where the dots stand for derivatives in $z$, and the effective masses are: 
$M_{\phi} = -{\partial P}/{\partial \beta}$, and
$M_{\psi} = M_{\theta} = (\cos^{2}{\theta_{-}} - \cos^{2}{\theta_{+}}) 
P$.  First, we consider a reduced model assuming an additional symmetry,  
$\theta _{+} = \pi/2$.  Then, the resulting system has 
{\em stable stationary points}. Solving this reduced system numerically we 
observe linear and even strong {\em nonlinear} oscillations near the 
stable minima. In general, 
the period of these oscillations in $R$ is different from the periods 
of $\theta_{-}$ and $\psi_{-}$. A stable stationary point corresponds to a 
smooth spiraling of the solitons.

\begin{figure}
\setlength{\epsfxsize}{7.2cm}
\centerline{\epsfbox{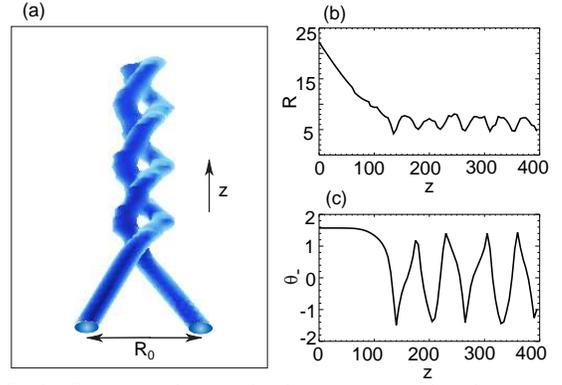}}
\caption{Stable soliton spiraling observed in direct modeling of Eqs. (1) 
for $\beta = -0.5$, $R_0 = \sqrt{500}$, $s = 10$, $V_0 = 0.2$, and $\theta 
_{-} = \theta _{+} = \pi/2$. (a) 3D view; (b) small oscillations of the 
relative distance between solitons; (c) large-amplitude oscillations for 
$\theta_{-}$ (quasi-periodic power exchange).}
\vspace{-0.4cm}
\label{fig1}
\end{figure}

\begin{figure}
\setlength{\epsfxsize}{7.2cm}
\centerline{\epsfbox{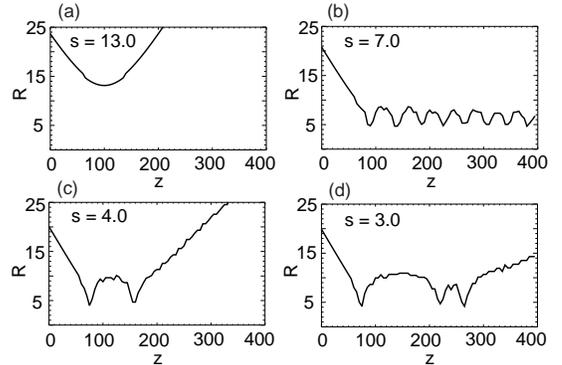}}
\caption{Examples of different dynamical regimes of the soliton 
interaction obtained by direct modeling of Eqs. (1) for $\beta = -0.5$, 
$V_0 = 0.2$,  and $\theta _{-} = \theta _{+} = \pi/2$. The initial 
separation is defined as $R_0 = \sqrt{400 + s^{2}}$. (a) Weak soliton 
interaction (no spiraling); (b) stable spiraling; (c)(d) unstable 
spiraling (decay of spiraling configuration via resonances).}
\vspace{-0.4cm}
\label{fig2}
\end{figure}

However, the analysis of the full dynamical system (\ref{full_system}) 
brings a surprise:  stable stationary points are absent. The main
reason for this is the negativeness of the effective mass $M_\phi$ in  
Eqs.~(\ref{full_system}), which is a typical destabilisation mechanism for 
any coherent soliton interaction
\cite{Vika}. However, numeriacl simulations show that {\em stable 
dynamical spiraling} is still possible. To understand the physical 
mechanism of such a {\em dynamical stabilisation},  we analyse 
the effective interaction potential (\ref{Inter}). Although $U_{\rm coh} > 
U_{\rm incoh}$ even for small $R$, large-scale periodic quarter-period  
out-of-phase oscillations in $\theta_{-}$ and 
$\psi_{-}$ can {\em significantly suppress} the effective value of the 
$U_{\rm coh}$ term, thus lowering its maximum value by a factor of $5$ or 
more. As a result, the incoherent attraction dominates and solitons become 
trapped in a spiraling configuration with oscillations near some $R_{\rm 
min}$ and {\em large-scale quasi-periodic oscillations} in both 
$\theta_{-}$ and $\psi_{-}$ (see Fig. 1).

Solving  Eqs. (\ref{norm_eqs}) [and also (\ref{full_system})] numerically, 
we confirm the mechanism of the dynamical soliton spiraling. In summary, 
our theory and numerics show that

(i) Trapping of two beams in a stable spiraling is possible for a large 
range of parameters [examples are shown, e.g., in Fig. 1, for $s=10$,  and 
Fig. 2(b), for $s=7$].

(ii) Initially {\em mutually incoherent} colliding solitons [{\em i.e.}, 
$\theta_{\mp}(0) = \pi/2$] become {\em partially
coherent} due to a periodic power exchange between their
components. Moreover, stable spiraling is always accompanied by a 
large-scale periodic power exchange.

(iii) Initially introduced partial coherence between interacting
solitons ({\em seed mutual coherence}) can result in repulsion of 
out-of-phase solitons and fusion of in-phase solitons, preventing 
spiraling.In this sense, modifying the initial mutual coherence can easily 
transform stable spiraling into repulsion (``escape'') or fusion.

(iv) For smaller $s$ and also for some values of $s$ where the spiralling 
and power-exchange frequencies become commensurable, the soliton spiraling 
is not possible [see Figs. 2(c,d)]. A series of `resonance windows', 
similar to those discovered for 2D soliton interactions \cite{camp}, are 
observed.  For such values of $s$, oscillations in $R$ are stronger and 
$U_{\rm coh}$ can become dominant (even being effectively suppressed), 
thus leading to a decay of spiraling.

To verify our theory, we perform a series of experiments. The experiments 
are carried out using the photorefractive screening nonlinearity 
\cite{Ming,one_dim}.
In essence, the photorefractive nonlinearity is anisotropic 
\cite{Crosig}, which makes it non-ideal to test our model. However, many 
experimental results suggest that for a large range of parameters, the 
anisotropy is fairly small:  isolated 3D solitons
 are almost fully circular \cite{Shih}, and planar collisions between 3D 
coherent solitons are almost fully isotropic \cite{Meng}, except for a 
special case, e.g., when the collision
plane is normal to the c-axis of the crystal and for  a particular initial 
distance between the solitons \cite{Krolik}. In this
respect, even though the photorefractive  nonlinearity is not 
isotropic in 3D, one can still employ it  to qualitatively  study the 
predictions of our theory. We therefore extrapolate the known analytic 
results for 2D photorefractive screening solitons \cite{one_dim}, which 
were all confirmed experimentally \cite{Kos}, to 3D which concurs with 
Eqs. (\ref{norm_eqs}).
\begin{figure}
\setlength{\epsfxsize}{6.7cm}
\centerline{\epsfbox{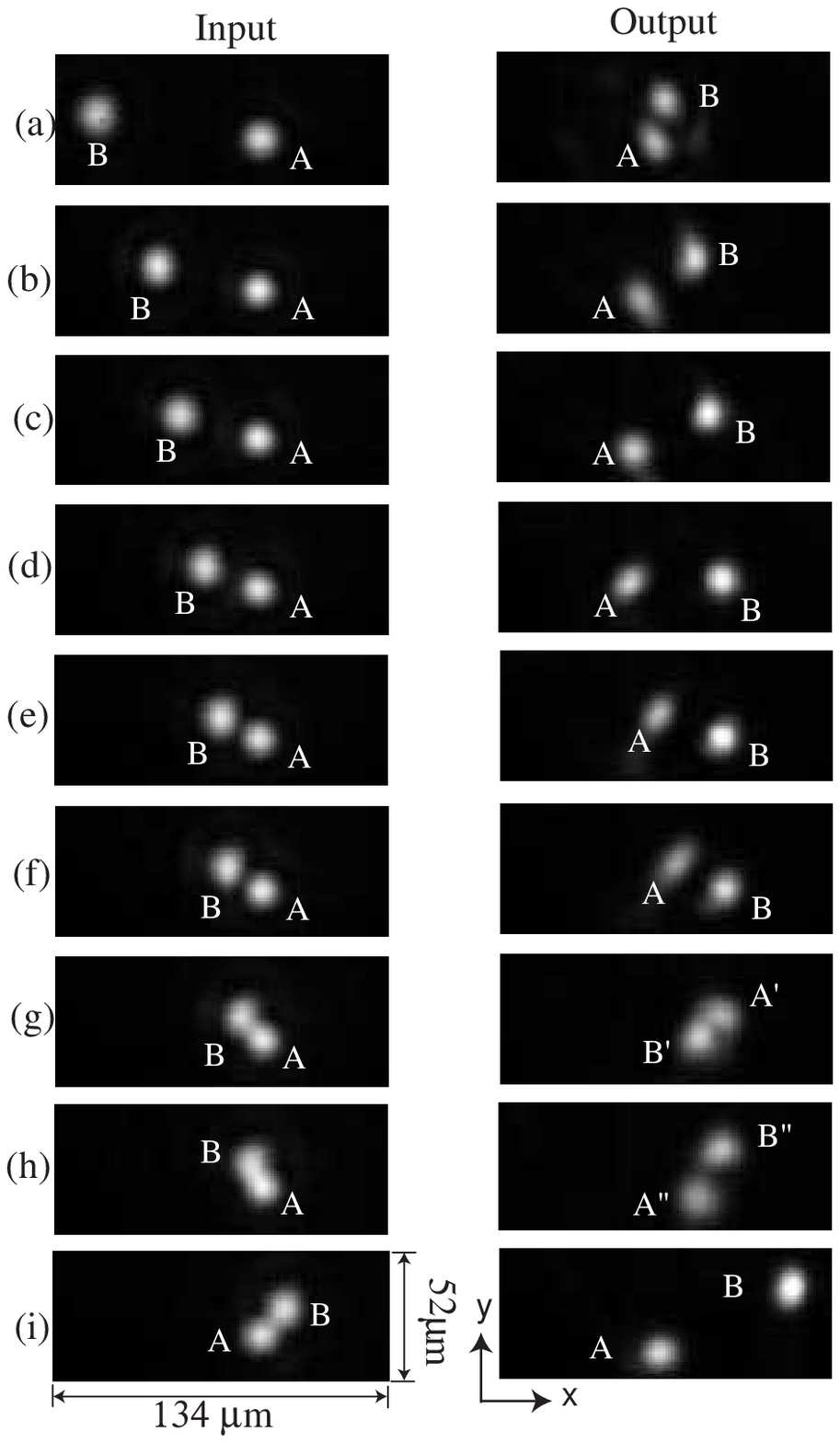}}
\vspace{0.2cm}
\caption{ (a)-(i) Collision of two initially mutually incoherent
solitons with different values of the impact parameter.}
\vspace{-0.4cm}
\label{fig3}
\end{figure}
\begin{figure}
\setlength{\epsfxsize}{7.1cm}
\centerline{\epsfbox{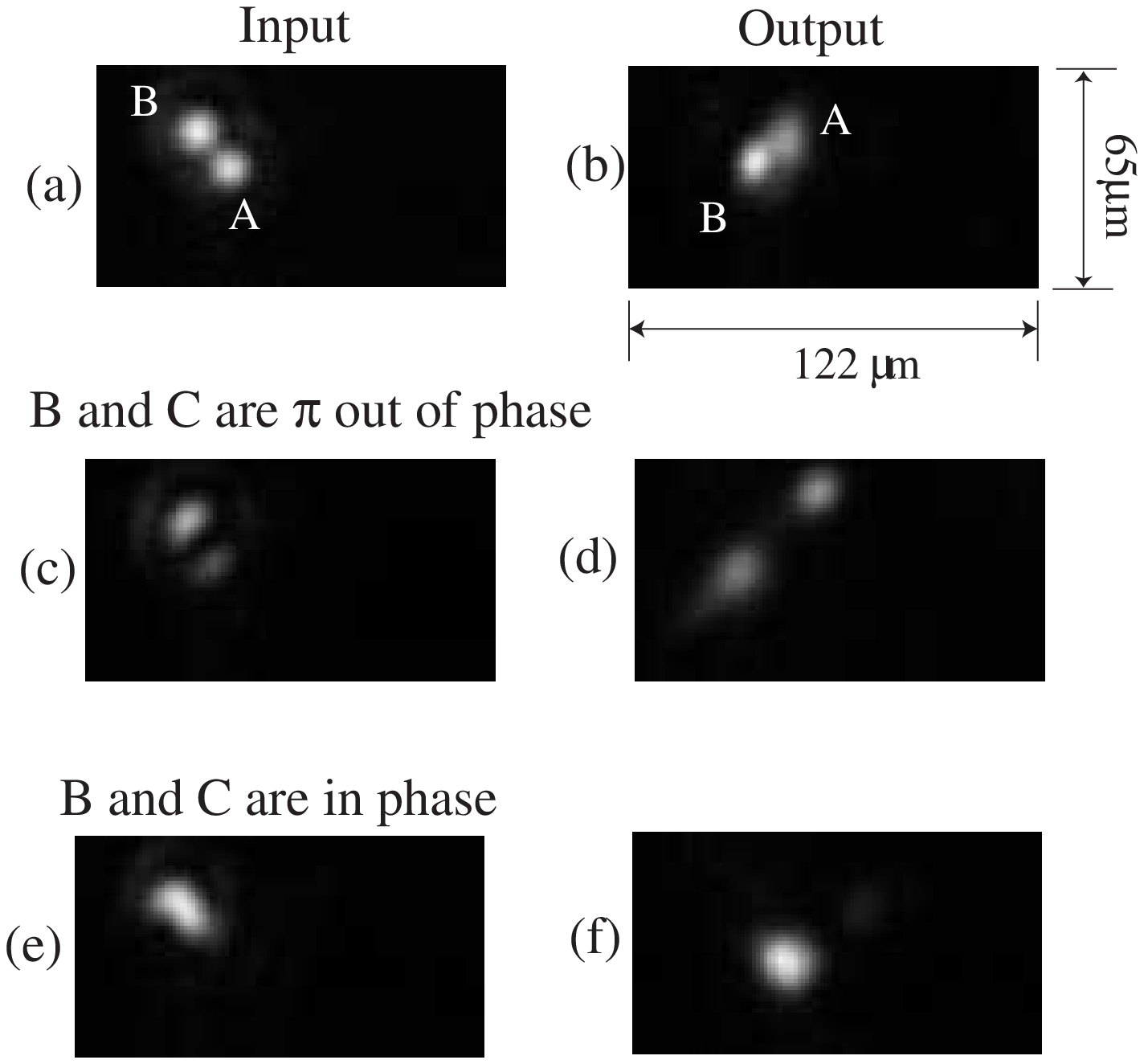}}
\vspace{0.2cm}
\caption{ (a),(b) Stable soliton spiraling of initially incoherent beams. 
(c),(d) A ``seed coherence'' beam C, which is $\pi$ out of phase with B, 
prevents the spiraling; (e),(f) When the beam C is in phase with B, it 
causes all three beams to fuse. }
\vspace{-0.4cm}
\label{fig4}
\end{figure}
The experimental setup is similar to that of Ref. \cite{Ming}.
Two soliton beams A and B of wavelength 488 nm, with power in the order of 
$\mu$W, and radii of 12$\mu$m FWHM are launched into an SBN crystal  whose 
electro-optic coefficient is 278 pm/V and the length is 6.5 mm. The 
initial $y$ coordinate of B is 9 $\mu$m higher than that of A, and B is 
launched with its initial trajectory inclined (relative to that of A)  by 
an angle of 0.01 radians
in the $x$ direction and 0.0012 radians in the $y$ direction. The 
intensity ratio between the soliton peak and the background illumination,  
$|u|^2_{bg} = |w|^2_{bg}$, is about 5.  A field of 4.2 kV/cm is applied  
against the the crystalline c-axis to generate the solitons. The impact 
parameter is adjusted by shifting the initial $x$ coordinate of B at the 
input, while all other initial conditions are kept unchanged. When the 
separation in the $x$ coordinate is larger than 31$\mu$m as shown in Fig. 
3(a)-3(c), solitons A and B barely interact [compare with Fig. 2(a), which 
shows a passing of two solitons as if they do not interact  at all]. As 
the impact parameter is reduced by shifting B closer to A, as shown in 
Fig. 3(d)-3(f), A and B's trajectories are bent due to the attraction 
force between them, and the amount of bending (scattering) is dependent on 
the impact parameter. This mimics a classical particle scattering 
experiment. We distinguish A from B and measure the power exchange by 
monitoring the output within a time window much shorter than the response 
time of the SBN crystal (1 sec) after A or B is blocked. The measured 
power exchange is smaller than 1\%  in Fig. 3(a)-3(f).

When we further reduce the separation in the $x$ coordinate to 9$\mu$m 
[Fig. 3(g)], the two solitons rotate around each other [cf. Fig. 1 and 
Fig. 2(b)].  We find that 60\% of A and 46\% of
B at the input go to A$^{\prime}$(at the output) and the rest goes to
B$^{\prime}$.  This power exchange is what we have called {\em induced 
coherence}. We also find that a small variation in B's initial position or 
trajectory which does not change the rotation angle of beam trajectories 
by much,  can cause the fraction of the exchanged power to vary 
considerably [compare with Fig. 1(c)].  In some spiraling cases, as low as 
a 5\% level of power exchange has been measured at the output of the 
crystal. In a similar spiraling experiment, but with different initial 
trajectories, we find
that the power exchange also depends on the intensity ratio. 
That is, the level of saturation of the nonlinearity. In that
experiment, 17\% power exchange is measured when solitons are generated  
with the intensity ratio of 12 and only 2\% for the intensity ratio of 4.
 
We then reduce the $x$ separation further to 4 $\mu$m [Fig. 3(h)],  and 
find that A and B interact strongly, but the spiraling seems to be 
unstable [compare to the numerical result shown in Fig. 2(c,d)]. Finally, 
when B is launched with its initial position beyond A [Fig. 
3(i)], they simply escape from each other.

In order to study how the initial partial coherence affects the soliton 
interaction, we introduce at the input a ``seed coherence'' beam C which 
is coherent with B but overlaps entirely and copropagates with A. When C 
is added, the intensity of A is reduced to make the total intensity (A+C) 
equal to that of B.
The relative phase between C and B is adjusted with a tilted piece of 
glass.  Before C is launched, we make sure the initial
conditions of A and B generate a spiraling pair [Figs. 4(a,b)]. When C is 
first adjusted to be out of phase with B [indicated by the dark notch 
between them at the input, Fig. 4(c)], B and A+C cannot spiral but just 
escape from each other, as shown in Fig. 4(d), although the power in  C is 
only about 28\% of A+C. When C is in phase with B, as shown in Fig. 4(e) 
(each intensity of A or C is 50\% of B), A, B and C fuse into one beam 
[Fig. 4(f)]. These experimental results agree with our theory,  
emphasizing the fact that seed coherence can be used to control the 
interaction outcome:  spiraling, ``escape'', or fusion.

In conclusion, we have analysed fully 3D interaction and spiraling of 
spatial solitons in an isotropic saturable bulk medium. The analysis, 
numerical simulations, and a series of experiments have revealed the 
important physical mechanism of the stable spiraling: a periodic power 
exchange between the interacting beams via induced coherence. Our results 
and conclusions are expected to hold for other types of 
(even anisotropic and nonsaturable) nonlinearity that depends on the total 
beam power and supports stable self-trapped beams in a bulk.

A.V.B and Yu.S.K. acknowledge support from the Australian Research Council 
and the Australian Photonics Cooperative Research Centre.  M-F.S. and M.S. 
were supported by the US Army Research Office.

\end{multicols}
\end{document}